\documentclass[twocolumn]{aastex62}
\usepackage{natbib}

\received{}
\revised{}
\accepted{May 20, 2019}

\submitjournal{ApJL}

\shorttitle{}
\shortauthors{}
\begin{document}

\title{THE GLOBULAR CLUSTER ORIGIN OF THE MILKY WAY OUTER BULGE: EVIDENCE FROM SODIUM BIMODALITY}

\correspondingauthor{Young-Wook Lee, Jenny J. Kim}
\email{ywlee2@yonsei.ac.kr, kim@uni-heidelberg.de}

\author[0000-0002-2210-1238]{Young-Wook Lee}
\affil{Center for Galaxy Evolution Research \& Department of Astronomy, Yonsei University, Seoul 03722, Republic of Korea}

\author[0000-0002-0432-6847]{Jenny J. Kim}
\affil{Center for Galaxy Evolution Research \& Department of Astronomy, Yonsei University, Seoul 03722, Republic of Korea}

\author[0000-0002-8878-3315]{Christian I. Johnson}
\affil{Center for Astrophysics $|$ Harvard \& Smithsonian, 60 Garden Street, Cambridge, MA 02138, USA}

\author[0000-0001-6812-4542]{Chul Chung}
\affil{Center for Galaxy Evolution Research \& Department of Astronomy, Yonsei University, Seoul 03722, Republic of Korea}

\author{Sohee Jang}
\affil{Center for Galaxy Evolution Research \& Department of Astronomy, Yonsei University, Seoul 03722, Republic of Korea}

\author[0000-0001-7277-7175]{Dongwook Lim}
\affil{Center for Galaxy Evolution Research \& Department of Astronomy, Yonsei University, Seoul 03722, Republic of Korea}

\author[0000-0002-5261-5803]{Yijung Kang}
\affil{Center for Galaxy Evolution Research \& Department of Astronomy, Yonsei University, Seoul 03722, Republic of Korea}

\begin{abstract}
Recent investigations of the double red clump in the color-magnitude diagram of the Milky Way bulge cast serious doubts on the structure and formation origin of the outer bulge. Unlike previous interpretation based on an X-shaped bulge, stellar evolution models and CN-band observations have suggested that this feature is another manifestation of the multiple stellar population phenomenon observed in globular clusters (GCs). This new scenario requires a significant fraction of the outer bulge stars with chemical patterns uniquely observed in GCs. Here we show from homogeneous high-quality spectroscopic data that the red giant branch stars in the outer bulge ($> 5.5^{\circ}$ from the Galactic center) are clearly divided into two groups according to Na abundance in the [Na/Fe] $-$ [Fe/H] plane. The Na-rich stars are also enhanced in Al, while the differences in O and Mg are not observed between the two Na groups. The population ratio and the Na and Al differences between the two groups are also comparable with those observed in metal-rich GCs. The only plausible explanation for these chemical patterns and characteristics appears to be that the outer bulge was mostly assembled from disrupted proto-GCs in the early history of the Milky Way.
\end{abstract}


\section{Introduction} \label{sec:intro}
\begin{figure*}
\includegraphics[scale=1.2]{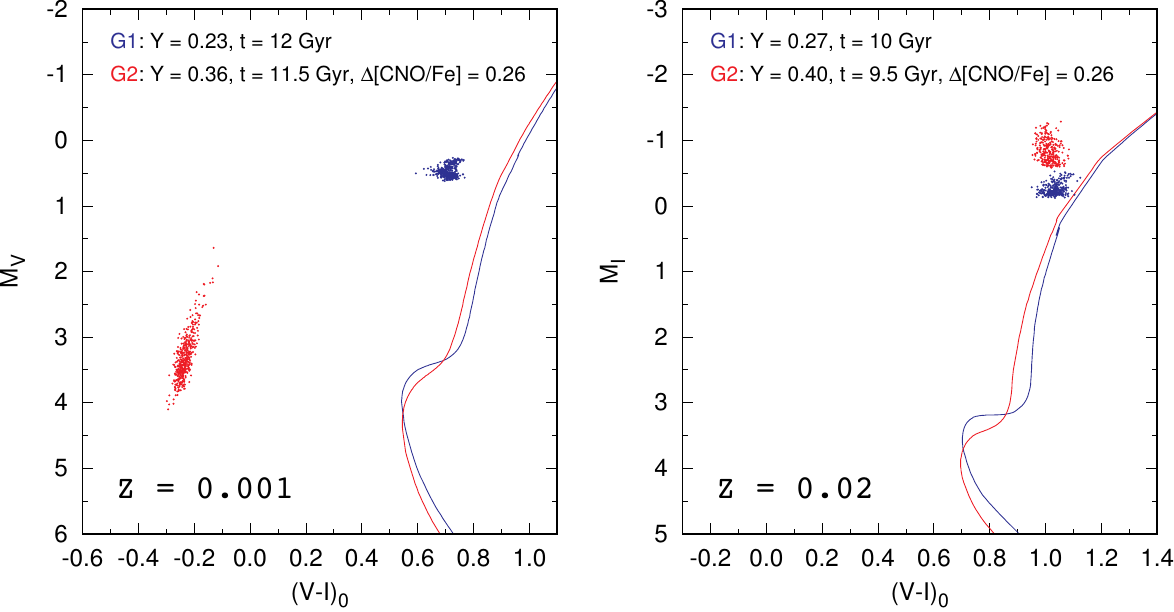}
\centering
\caption{Stellar population models for G1 and G2 stars at two different metallicity regimes. These models, based on the Yonsei$-$Yale ($Y^{2}$) HB evolutionary tracks and isochrones, are to illustrate the variation of HB morphology with metallicity (see \citealt{lee2015} {and \citealt{joo2017}} for the details of model construction {and for similar models with no [CNO/Fe] variation}). In the metal-poor ($Z = 0.001$) GCs, super-He-rich G2 stars are placed at bluer HB; however, they are placed at bright RC in the metal-rich ($Z = 0.02$) regime like the bulge. The $\Delta$${\rm Y(G2 - G1)} = 0.13$ is fixed in the two panels {for the illustration purpose. $\Delta {\rm Y}\simeq 0.13$ and $\Delta {\rm [CNO/Fe]} \simeq 0.26$ are predicted by \citet{kim2018} chemical evolution models for solar metallicity GCs.}
\label{fig1}}
\end{figure*}

The Milky Way bulge appears to have properties of both ``{classical bulge}" (CB) with merger origin and ``{pseudo bulge}" composed of the bar/disk populations, although it is debated whether the contribution from CB is negligible or significant \citep{nataf2017, zoccali2018}. Since the inner bulge is dominated by a prominent bar component \citep{blitz1991, bland-hawthorn2016}, the presence of CB component would be best traced at the outer bulge. However, the discovery of the double red clump \citep[RC;][]{mcwilliam2010, nataf2010} in high-latitude ($|b|> 5.5^{\circ}$) field of the bulge led to the consensus that the bar is extended even to the outer bulge, as it was initially thought to be a result of the distance difference between the two arms of a giant X-shaped structure originated from the bar instability \citep{mcwilliam2010, li2012, wegg2013}. A drastically different interpretation has been suggested for the double RC phenomenon, however, based on the multiple stellar populations observed in globular clusters \citep[GCs;][]{lee2015, lee2016, joo2017}. It is now well-established that most GCs host second- and later-generation ({G2+}) stars enriched in He, Na, Al, and N, together with first-generation (G1) stars having normal chemical composition. This is a unique phenomenon only observed in GCs where the chemical evolution is dictated by asymptotic giant branch (AGB) stars and winds of massive stars (WMS) with negligible contribution from supernovae \citep{gratton2012, renzini2015, kim2018}. In metal-poor GCs, super-He-rich {G2+} stars are usually placed on the blue horizontal-branch (HB), however, in the metal-rich regime, like the bulge, they are naturally placed on the bright RC, producing the double RC in the color-magnitude diagram (CMD; see Figure~\ref{fig1}), as is observed in a metal-rich bulge GC Terzan 5 \citep{ferraro2009, joo2017}.

Recently, low-resolution spectroscopy has confirmed that the stars in the two RC regimes show a significant difference in CN-band strength which is comparable to those observed in GCs between N-rich {G2+} and N-normal G1 stars, supporting the multiple population interpretation \citep{lee2018}. According to this new scenario, most stars in the outer bulge were provided by disrupted proto-GCs, and therefore the unique chemical patterns of multiple populations in GCs must be observed not only among RC stars but also among red giant branch (RGB) stars in the outer bulge. In order to scrutinize this prediction, we have revisited the spectroscopic data available in the literature for RGB stars in the outer bulge of the Milky Way and compared them with our chemical evolution models for metal-rich GCs.

\section{Spectroscopic data} \label{sec:2}
\begin{figure*}
\includegraphics[scale=.45]{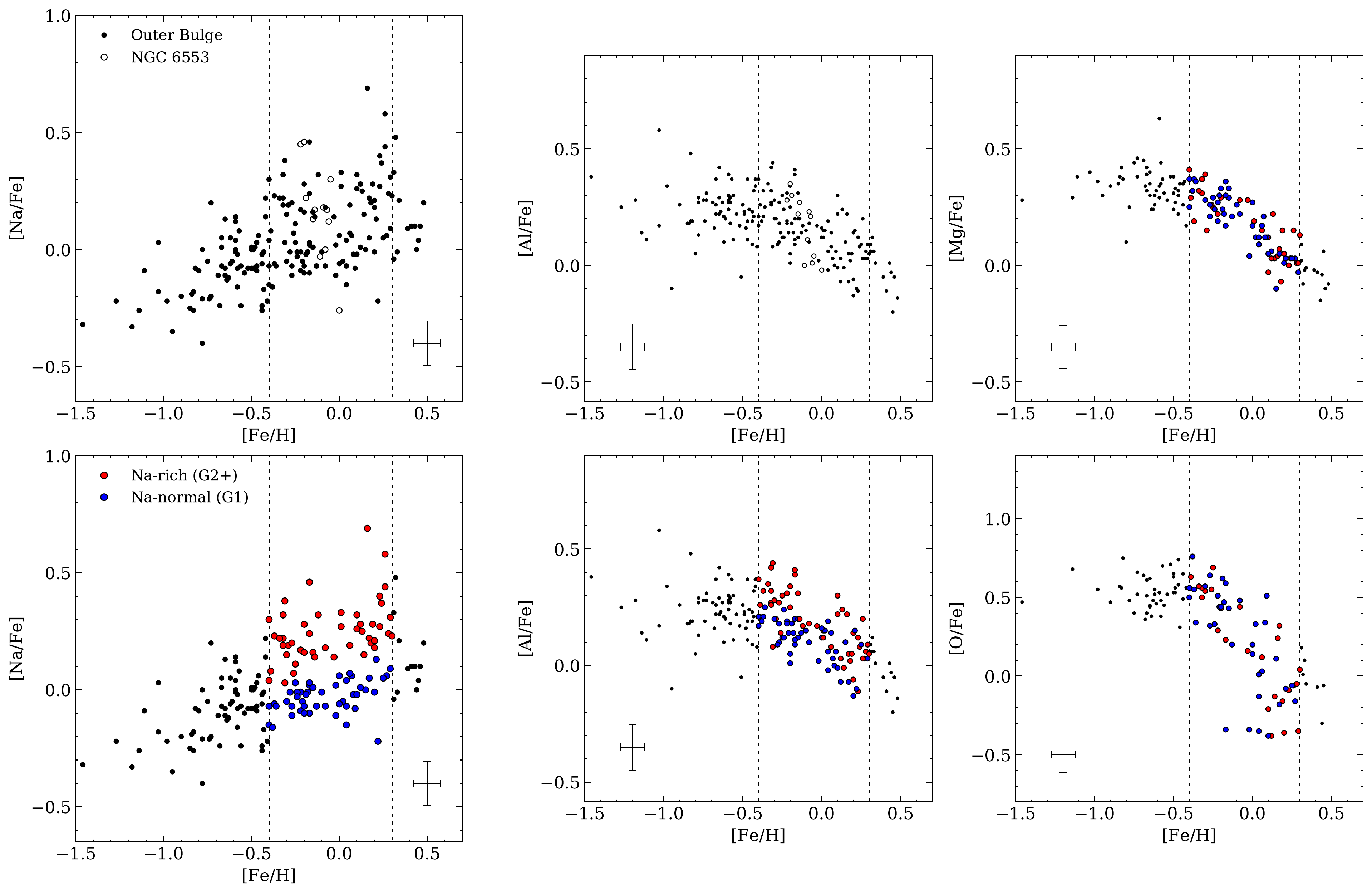}
\centering
\caption{Na, Al, O, and Mg abundances of RGB stars in the outer bulge of the Milky Way as a function of Fe abundance. In the left two panels, two sequences of stars are most clearly seen in the relatively metal-rich regime ($-0.4 < [Fe/H] < 0.3$; vertical dashed lines) where the double RC is observed. Typical measurement uncertainties ($\pm 1 \sigma$) shown are random errors.
\label{fig2}}
\end{figure*}

Data homogeneity and quality are essential in this study because the expected differences in [Na/Fe] and [Al/Fe] between {G2+} and G1 stars in metal-rich GCs are predicted to be only $\sim$0.3 and $\sim$0.1 dex, respectively (\citealt{kim2018}; see also Figure~\ref{fig4} below). Therefore, for the sake of maintaining data homogeneity while ensuring a sufficient sample size, we have employed and combined two high-quality data sets obtained by the same investigator \citep{johnson2012, johnson2014} in three bulge fields centered at $(l, b) = (-1^{\circ}, -8.5^{\circ}), (0^{\circ}, -12^{\circ})$, and $(5.3^{\circ}, -3^{\circ})$. The sample stars are in the outer bulge ($>$ 5.5$^{\circ}$ from the Galactic center) and, in terms of the Galactocentric distance ($R_{gc}$), most of them are distributed roughly in $0.8 < R_{gc}< 1.8$ kpc. 

\begin{figure}
\includegraphics[scale=.5, trim=2cm 5.7cm 2cm 5.5cm, clip=True ]{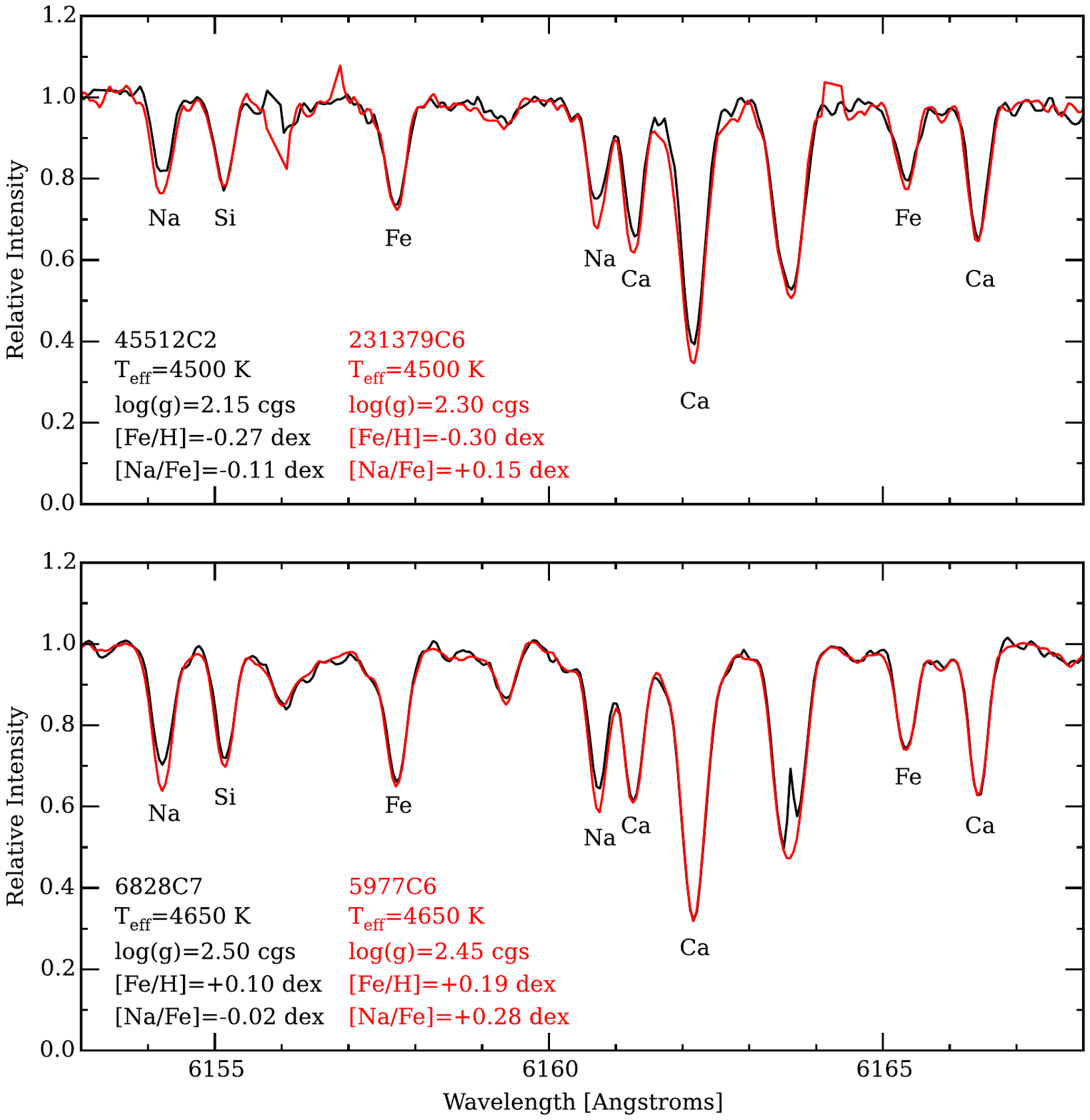}
\centering
\caption{Example of the spectra for the two stars in Na-rich and Na-normal groups. Two spectra, each representing stars in the Na-rich (red) and Na-normal (black) groups, are compared for two pairs of stars. They have a nearly identical temperature, gravity, and [Fe/H]. Note the difference in Na line strength between the two spectra.
\label{fig3}}
\end{figure}

Although the data were taken in different years and with different instruments, the Hydra ($b = -8.5^{\circ}$ field) and FLAMES ($b = -12^{\circ}$ and $-3^{\circ}$ fields) multi-fiber spectrographs produce similar resolving powers (R $ \approx 20,000$). The target stars span comparable ranges of effective temperature, surface gravity, and metallicity, and the spectra exhibit comparable signal-to-noise ratios ($\sim50 - 100$ per reduced pixel). As a result, the Na, Al, and Fe absorption line profiles of interest to the present work were equivalently resolved, and any stated abundance ratio differences between and within the data sets should only reflect random measurement uncertainties and intrinsic composition variations. Systematic abundance ratio differences are largely minimized by our selection of the data sets from the same investigator. Both studies used the same LTE line analysis code \citep[MOOG;][]{sneden1973}, equivalent width (Fe) and spectrum synthesis (Na, Al) techniques, the same absorption lines, and the same model atmosphere grids. The reported abundance ratios in both studies were also measured differentially relative to the standard giant star Arcturus, which should largely reduce systematic effects due to model atmosphere deficiencies, departures from thermodynamic equilibrium, and departures from the standard 1D plane parallel atmosphere assumption. We expect systematic zero point differences do not exceed $\sim$0.03 dex in magnitude.

\section{Sodium bimodality in the outer bulge} \label{sec:3}
\begin{figure*}
\includegraphics[scale=.8 ]{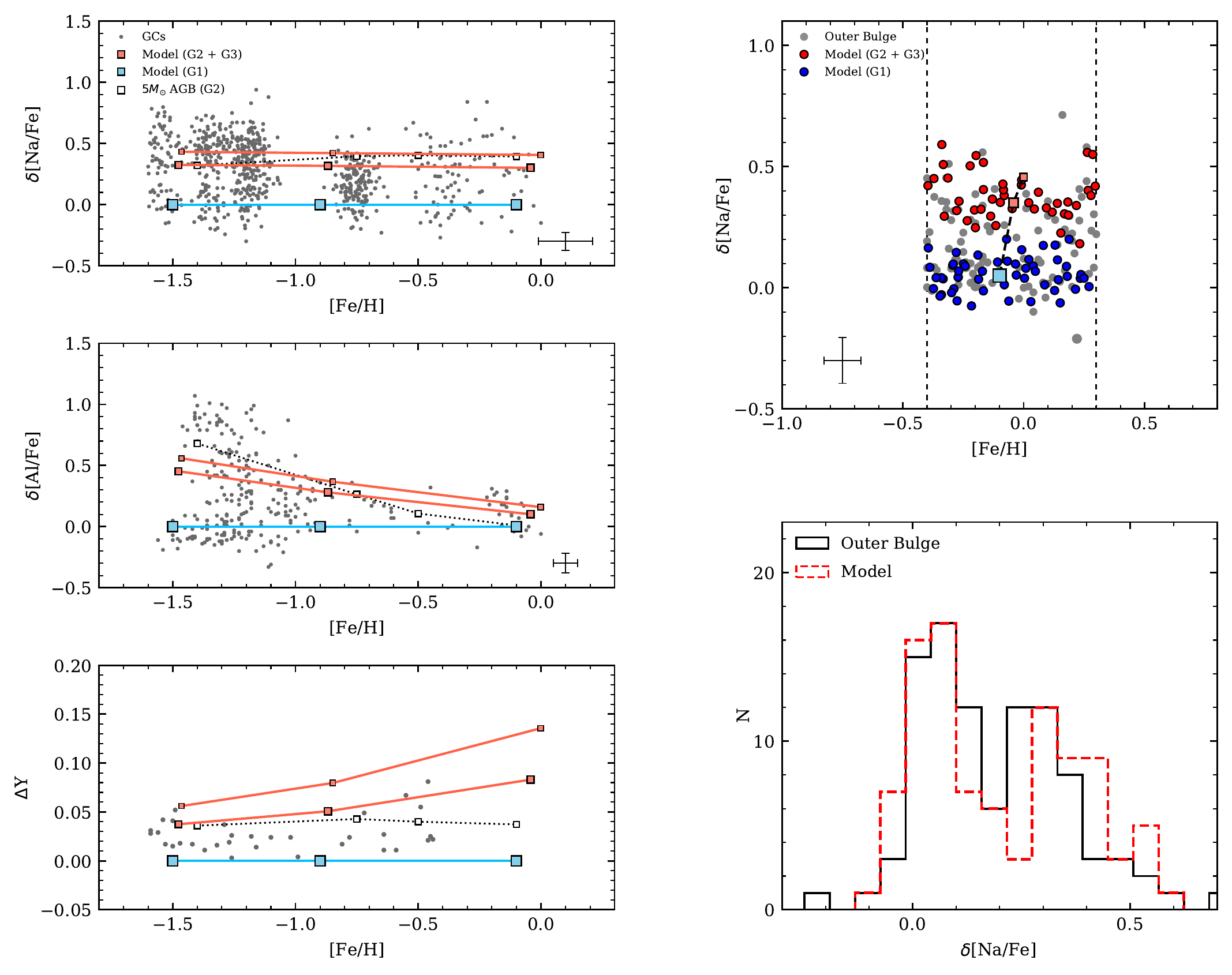}
\centering
\caption{
In the left panels, chemical evolution models \citep{kim2018} for G1, G2, and G3 (third generation) are compared with the observational data for GCs. Also compared is a simple model in AGB yield only scenario (40\% of $5 M_{\odot}$ AGB ejecta diluted with 60\% of pristine gas; yields from \citealt{2016MNRAS.458.2122D}, and references therein). Note that $\delta {\rm [Al/Fe]}$ between G2 and G1 decreases rapidly with increasing ${\rm [Fe/H]}$ and, therefore, it is predicted to be much smaller than $\delta {\rm [Na/Fe]}$ at metal-rich regime. Observational data compared are for NGC 104, 362, 1904, 5904, 6121, 6139, 6440, 6441, 6388, 6528, and 6553 for Na, and for Palomar 6, NGC 5272, 5904, 6171, 6528, 6553, and 6838 for Al \citep[][and references therein]{carretta2016, munoz2018}. The predicted variations of He abundance ($\Delta {\rm Y}$) from the same models are also shown in the lower panel with the values derived by \citet{2018MNRAS.481.5098M}. Note that $\Delta {\rm Y}$'s estimated from CMD are somewhat uncertain \citep[see Figure 18 of][]{2018MNRAS.481.5098M}, while those predicted from chemical evolution models also vary with adopted parameters and yield \citep[see Figure 12 of][]{kim2018}. In the right panels, the distribution of $\delta {\rm [Na/Fe]}$ predicted from our metal-rich models (convolved with observational errors) is compared with the observed data for the outer bulge.
\label{fig4}}
\end{figure*}

Figure~\ref{fig2} shows [Na/Fe], [Al/Fe], [O/Fe], and [Mg/Fe] abundances of RGB stars in these outer bulge fields plotted as a function of [Fe/H]. The most striking feature is a Na-bimodality, with Na-rich and Na-normal sequences, which is evident in the [Na/Fe] $-$ [Fe/H] diagrams in the left two panels.\footnote{All three fields show a similar bimodality or spread in Na although with some stochastic variations expected from small sample size per each field, and therefore the Na-bimodality is most clearly seen when they are combined into one large sample.} This bimodal Na distribution is most clearly seen in the relatively metal-rich regime ($-0.4 <$ [Fe/H] $< 0.3$) where the double RC is observed \citep{ness2012, uttenthaler2012}. Figure~\ref{fig3} compares two typical examples of spectra, each representing stars in the Na-rich and Na-normal groups, which shows that the Na line strength variations are real. In order to further investigate this finding in terms of GC connection, in the lower left panel of Figure~\ref{fig2}, Na-rich and Na-normal stars are highlighted in different color coding in the referred metallicity range. The same color coding is then repeated in other panels for Al, O, and Mg abundances. If the bimodal Na distribution is due to the putative {G2+} and G1 stars originated in metal-rich GCs, Na-rich stars should also be observed to be Al-rich although the difference in [Al/Fe] between the two groups is predicted to be substantially smaller than that for [Na/Fe] (see Figure~\ref{fig4}).\footnote{The yields for metal-rich fast rotating massive stars are not available in the literature, but our models \citep{kim2018} constructed with only WMS yields (\citealt{2018ApJS..237...13L} with moderate rotation) also show a similarly strong variation of $\delta{\rm [Al/Fe]}$ with metallicity.} This is clearly seen in the lower middle panel. Furthermore, unlike metal-poor GCs, {G2+} stars in metal-rich GCs are not significantly depleted in O and Mg abundances ($\delta$[O/Fe] $\approx$ 0.04 dex; \citealt{tang2017, munoz2018, kim2018}),  and therefore, to within the observational errors, we would not expect differences in O and Mg abundances between {G2+} and G1 stars. This prediction is also consistent with the absence of apparent differences in [O/Fe] and [Mg/Fe] between the Na-rich and Na-normal groups in the right two panels. Similarly, no obvious chemical dichotomy is observed in other elements measured including Si, Ca, Ni, Co, and Cr. Therefore, all of these chemical patterns observed in the outer bulge are exactly consistent with the behavior we would expect from {G2+} and G1 stars originated in metal-rich GCs.

To test the statistical significance of the bimodal Na distribution of our bulge sample, in Figure~\ref{fig5}, the relative differences of Na abundance from the lower boundary line are plotted in the same metallicity range of $-0.4 <$ [Fe/H] $< 0.3$. The Gaussian Mixture Modeling \citep[GMM;][]{muratov2010} test confirms that the distribution is bimodal at a very high level of significance ($p < 0.001$ with the peak separation ratio $D = 3.0$). For comparison, we have also plotted the same diagrams for the disk stars from which we can see that the Na-bimodality is not observed in the disk stars as they show only single G1 characteristic.\footnote{Note that this difference cannot be attributed to the different evolutionary stages of the measured stars in the outer bulge (RGB stars) and in the disk (main-sequence stars) as similar spreads in Na abundance are observed among main-sequence stars in GCs \citep{carretta2016}.}  This shows that the outer bulge stars are unlikely to have been provided from the disk as is believed in the X-shaped bulge scenario for the double RC phenomenon. Although the sample size is small, the data obtained at the inner bulge \citep[$|b| < 5^{\circ}$;][]{bensby2013} show a similarly G1 dominated distribution but with some stars having a weak {G2+} characteristic.

Quantitatively, the differences in [Na/Fe] and [Al/Fe] between the two sequences of bulge stars are 0.27 $\pm$ 0.02 and 0.10 $\pm$ 0.02 dex, respectively, which are in good agreements with those between {G2+} and G1 stars observed in metal-rich GCs and those predicted from our chemical evolution models ($\delta$[Na/Fe] $\approx$ 0.3, $\delta$[Al/Fe] $\approx$ 0.1; see Figure~\ref{fig4}). The population ratio between Na-rich and Na-normal groups is 0.49:0.51, which is consistent with the ratio of the bright and faint RC stars observed in the CMD of the outer bulge fields \citep[0.52:0.48;][]{lee2018}. Interestingly, they are also comparable to the number ratio of {G2+} and G1 stars observed in GCs{, those with $M_V > -7.5$ in particular} \citep{milone2017}. The consistency of the abundance trends between the stars in metal-rich GCs and those in our bulge fields can also be directly confirmed from the data for a metal-rich bulge GC NGC 6553 obtained from the same observing run (see Figure~\ref{fig2}).

\section{Discussion} \label{sec:4}
\begin{figure*}
\includegraphics[scale=.7 ]{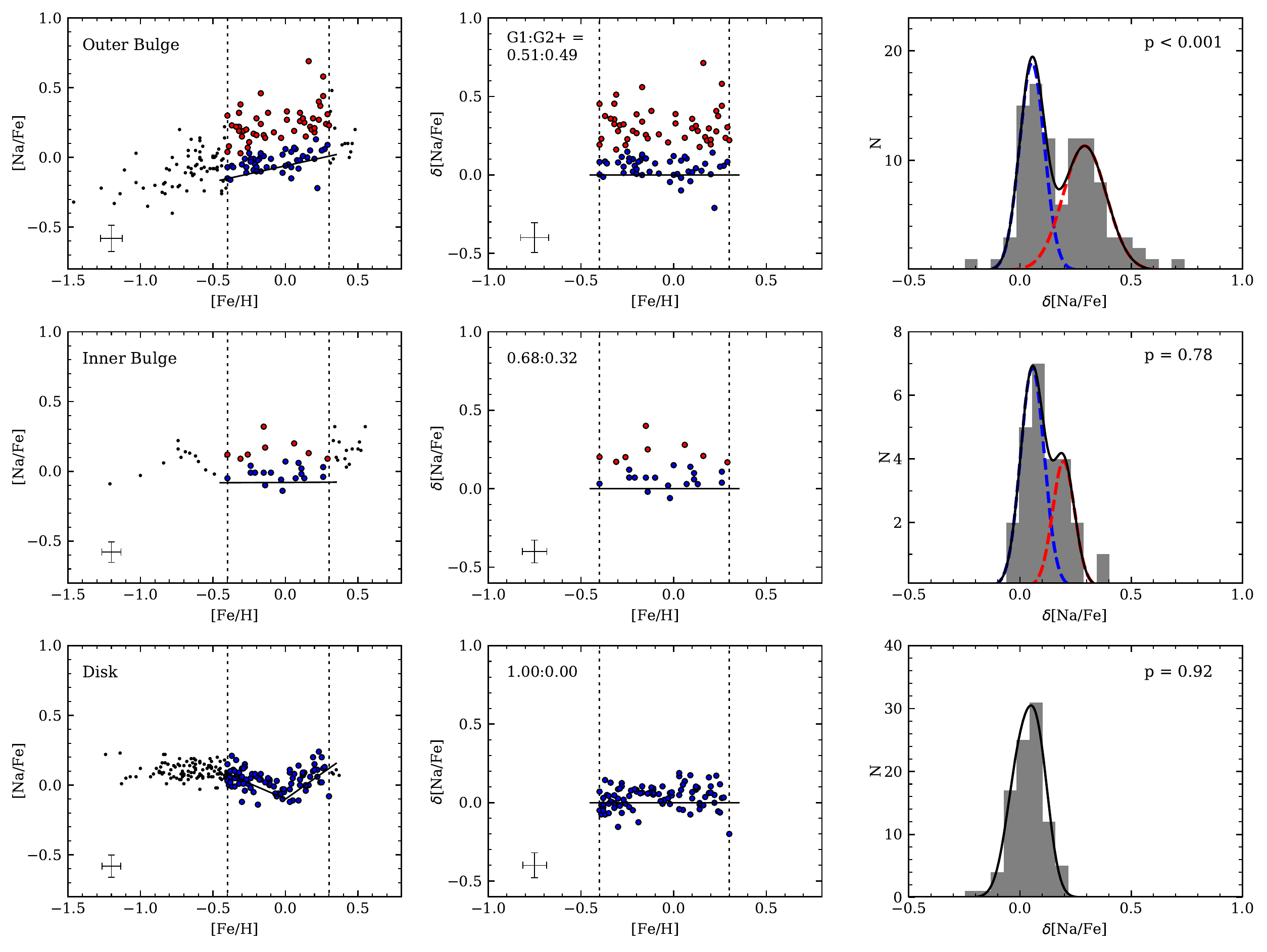}
\centering
\caption{Relative differences of [Na/Fe] with respect to the lower boundary lines are plotted. In the upper panels, our sample stars in the outer bulge show a large spread in $\delta$[Na/Fe] with a strong bimodality from the GMM test ($p < 0.001$; $D = 3.0$). In the lower panels, however, disk stars \citep{bensby2005, reddy2006} show no chemical dichotomy ($p = 0.92$) with a narrow spread in $\delta$[Na/Fe]. The middle panels are for the stars in the inner bulge at low-latitude fields \citep[$|$b$|$ $<$ 5$^{\circ}$;][]{bensby2013}, which show a similarly G1 dominated distribution but with some stars having a weak G2 characteristic.
\label{fig5}}
\end{figure*}

In summary, our sample of outer bulge stars shows (1) strong Na-bimodality, (2) unique chemical patterns displayed by Na, Al, O, and Mg only observed and predicted in metal-rich GCs, (3) the differences in Na and Al abundances between the putative {G2+} and G1 stars fully consistent with those observed and theoretically predicted in metal-rich GCs, and (4) the Na-rich/Na-normal population ratio which is comparable to those estimated from the bright and faint RC stars in the outer bulge fields and from {G2+} and G1 stars in GCs. All of these characteristics are only explained by {G2+} and G1 stars originated in GCs (or subsystems similar to GCs in terms of chemical enrichment), empirically and theoretically. It is now well established that the unique chemical abundance patterns associated with multiple populations in GCs are not observed and theoretically predicted in other systems and environments such as dwarf galaxies, open clusters, and the Milky Way disk \citep{gratton2012, renzini2015, bastian2018, hasselquist2017}. The observed Na-bimodality and the accompanying Al difference in our bulge sample cannot arise from the environments where the supernova ejecta is retained and dominating chemical enrichment, such as the present day bulge, because the bimodality is observed at given metallicity and no other chemical dichotomy is detected in other elements measured. Some Na enrichment is observed from the first dredge-up \citep{reddy2019}, but this effect is not relevant to our sample of bulge RGB stars because our Na-rich and Na-normal samples are all brighter than RGB bump and RC with similar distributions on CMD. We have also checked the radial velocity and Gaia proper motion and did not find any meaningful differences in kinematic properties between the two Na groups. Note that some difference in proper motion reported recently \citep{sanders2019} between the two RC regimes at $(l, b) = (0^{\circ}, -8^{\circ})$ was predicted by our composite bulge model \citep[see Figure 11 of][]{joo2017} because some {foreground} bar component is superimposed on the brighter RC regime in the luminosity function.

Although the present result is mainly for the outer bulge, the presence of some stars enhanced in Na and N even in the inner bulge \citep[middle panels of Figure~\ref{fig5} and][]{schiavon2017} indicates a composite nature of the Milky Way bulge with a prominent bar embedded in our ``{GC-origin bulge}". In this respect, it is interesting to note that, at given metallicity, a relatively large spread is observed for Na from the APOGEE inner bulge ($|b| < 5^{\circ}$) data \citep{zasowski2019}, although the measurement error is larger for Na in this survey performed at near-IR in the Northern hemisphere. Considering the population ratio of the putative {G2+} and G1 stars from our result, which is comparable to those observed in GCs, a significant population of stars in the outer bulge region would have assembled from disrupted proto-GCs in the early history of the Milky Way.\footnote{In order to explain the ``mass budget problem'' \citep{renzini2015}, some scenarios for the formation of multiple populations in GCs \citep[e.g.,][]{2010MNRAS.407..854D} assume that G1 stars were preferentially lost into the field. In the outer bulge fields, the population ratio of G2+ and G1 stars is already roughly comparable to those observed in GCs, which would suggest that the preferential loss of G1 stars would be far smaller than those required in these models. Note, however, that the mass budget problem is much alleviated in \citet{kim2018} model, and therefore this population ratio issue does not pose any serious problem in their scenario.} This should be compared to the result obtained from the chemical tagging based on CN and CH in the halo, where only $\sim$3\% of the stars show GC-like {G2+} characteristics \citep{martell2011}. This comparison suggests that GCs played a major role as building blocks in the outer bulge, while most stars in the outer halo were probably accreted from dwarf galaxies with little contribution from GCs. In this context, the apparent absence of Na-bimodality among metal-poor stars ([Fe/H] $< -0.4$) in Figure~\ref{fig2} might indicate that they could be halo population embedded in the bulge region{, although it might be simply due to a small sample size at metal-poor tail of the metallicity distribution function}.

In broad terms, our ``{GC-origin bulge}" is consistent with recent cosmological hydrodynamic simulations of galaxy formation \citep{kruijssen2015, pfeffer2017, renaud2017}, which predict that proto-GCs form in high pressure environments in proto-disk galaxies at high redshift, and as they merge, stars from disrupted GCs would end up at the bulge of a Milky Way-like galaxy. Our ``{GC-origin bulge}" is also compatible with the ``{clump-origin bulge}" \citep{noguchi1999, elmegreen2008, inoue2012}, which can reproduce both the observed rotation and boxy shape of the bulge. Clumps are observed in proto-galaxies at high redshift \citep{vanzella2017} and many GCs would form within a single clump as suggested by theoretical investigations \citep{bekki2017, elmegreen2018}. Therefore, as the clumps containing proto-GCs were migrated and sucked into the galactic center, G1 and {G2+} stars from disrupted GCs would again end up in the bulge as well in this scenario. Further spectroscopic observations of RC and RGB stars at various latitude and longitude fields of the Milky Way bulge would provide crucial information on the population mixtures at different sightlines towards the bulge. Eventually, the data to be obtained from this survey, together with the Gaia distances, would help us to figure out the 3D structure and more detailed assembly history of the Milky Way bulge, which could also shed new light on our understanding of stellar populations and formation history of early-type galaxies.

\acknowledgments
{We thank the anonymous referee for a number of helpful comments and suggestions.}
Support for this work was provided by the National Research Foundation of Korea (grants 2017R1A2B3002919, and 2017R1A5A1070354).


\end{document}